\documentclass[sigconf]{acmart}
\renewcommand\footnotetextcopyrightpermission[1]{} 
\usepackage{xspace}
\usepackage{graphicx}
\newcommand\myname{VIRAL\xspace}

\usepackage[dvipsnames]{xcolor}
\usepackage[table]{xcolor}
\usepackage{subcaption}
\usepackage{listings}
\usepackage[normalem]{ulem}

\usepackage{wrapfig}
\usepackage{amssymb}
\usepackage{amsmath}
\usepackage{bm}
\usepackage{caption}
\usepackage{multirow}
\usepackage{mathrsfs}
\usepackage{mathtools}
\usepackage{algorithm}
\usepackage{algpseudocode}
\usepackage{soul}
\usepackage{booktabs}

\AtBeginDocument{%
  }

\begin{document}

\title{Enhancing Multimodal Recommendations with Vision-Language Models and Information-Aware Fusion}

\author{Hai-Dang Kieu}
\affiliation{%
\institution{University of Technology Sydney}
\city{Sydney}
\country{Australia}
}

\email{haidang.kieu@student.uts.edu.au}

\author{Min Xu}
\affiliation{%
  \institution{University of Technology Sydney}
  \city{Sydney}
  \country{Australia}
}
\email{min.xu@uts.edu.au}

\author{Thanh Trung Huynh}
\affiliation{%
  \institution{VinUniversity}
  \city{Hanoi}
  \country{Vietnam}
}
\email{trung.ht@vinuni.edu.vn}

\author{Dung D. Le}
\affiliation{%
\institution{VinUniversity}
\city{Hanoi}
\country{Vietnam}
}
\email{dung.ld@vinuni.edu.vn}

\renewcommand{\shortauthors}{HKieu}

\begin{abstract}

Recent advances in multimodal recommendation (MMR) highlight the potential of integrating visual and textual content to enrich item representations. However, existing methods often rely on coarse visual features and naive fusion strategies, resulting in redundant or misaligned representations. From an information-theoretic perspective, effective fusion should balance unique, shared, and redundant modality information to preserve complementary cues. To this end, we propose \textbf{\myname}, a novel Vision–Language Information-aware Recommendation framework that enhances multimodal fusion through two components: (i) a VLM-based visual enrichment module that generates fine-grained, title-guided descriptions for semantically aligned image representations, and (ii) an information-aware fusion module inspired by Partial Information Decomposition (PID) to disentangle and integrate complementary signals. Experiments on three Amazon datasets show that VIRAL consistently outperforms strong multimodal baselines and substantially improves the contribution of visual features\footnote{Code is available at: https://github.com/dangkh/VLIF}.

\end{abstract}



\maketitle
\vspace{-6pt}
\section{Introduction}

In recent years, multimodal recommendation (MMR) has emerged as a promising approach that integrates diverse item-related content—such as images, text, and metadata—to enrich behavioral signals. These additional modalities provide rich semantic cues that complement behavioral signals. For example, product images often highlighting features that attract buyers and enhance recommendation accuracy.

\begin{figure}[h!]
    \centering
    \includegraphics[width=.6\linewidth]{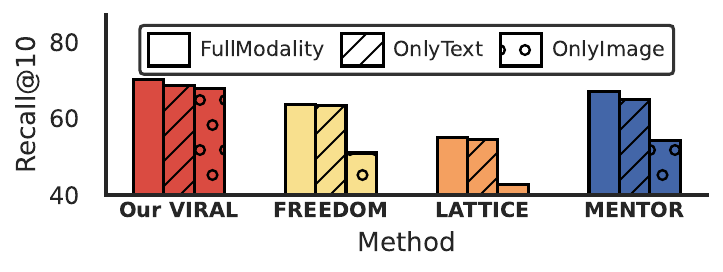}
    \includegraphics[width=.6\linewidth]{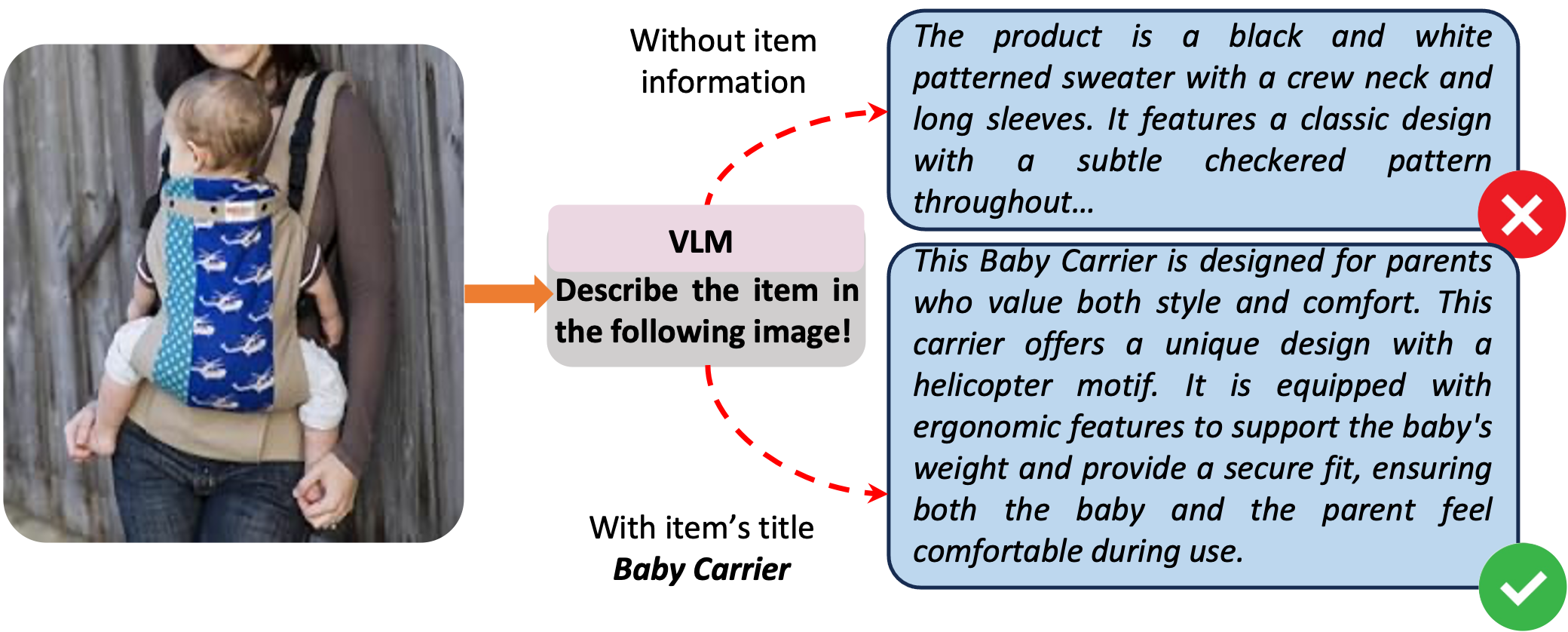} 
    \vspace{-0.2in}
    \caption{[Top] Our \myname outperforms recent SOTA multimodal models. [Bottom] Using VLM to generate visual description of item.}     
    \vspace{-0.25in}
    \label{fig:modality}
\end{figure}

A key challenge in MMR is effectively fusing heterogeneous modalities to construct expressive, semantically aligned representations~\cite{COHESION}. Despite notable progress, the visual modality contributes little to overall performance—removing image features typically causes only minor degradation, while textual signals remain dominant~\cite{FREEDOM, Mentor}. This occurs because product images are often optimized for marketing aesthetics rather than descriptive accuracy, which can mislead visual encoders and reduce semantic fidelity (see~\autoref{fig:modality}[Top]). These issues highlight the need to incorporate textual context to better guide visual understanding.

\textit{Vision–Language Models} (VLMs) generate semantically rich representations that capture fine-grained item details and mitigate the limitations of raw visual features~\cite{Rec-gpt4v, largeVisionEmpirical}. However, generic captioning often produces off-target content that fails to reflect true item semantics. As shown in ~\autoref{fig:modality}[Bottom], when an image contains multiple objects (e.g., a mother holding her daughter), a VLM prompted without item context may focus on irrelevant details such as clothing rather than the actual product (e.g., a baby carrier). Incorporating textual cues (e.g., the item title) enables more accurate, context-aware descriptions, underscoring the value of text-guided visual enrichment in multimodal recommendation.

Nonetheless, naive fusion strategies~\cite{DRAGON, Mentor} often treat modalities equally, which can introduce redundancy, conflicting cues, or an over-reliance on textual information—ultimately degrading representation quality. To overcome these challenges, \textit{Partial Information Decomposition} (PID)~\cite{quantifying} offers a principled framework for quantifying modality interactions. By isolating complementary and consistent information across modalities, PID-based approaches facilitate more effective fusion and lead to more robust, semantically aligned multimodal representations for recommendation tasks.

To address these issues, we propose \textbf{V}ision–Language \textbf{I}nformation-aware \textbf{R}ecommendation with \textbf{A}daptive \textbf{L}earning (\textbf{\myname}), a novel framework for multimodal recommendation. Instead of directly relying on raw image features, \myname leverages VLMs guided by item information (e.g., titles) to generate fine-grained, context-aware descriptions that capture essential item semantics. Furthermore, an information-aware fusion module inspired by PID is designed to quantify synergistic and redundant interactions between modalities, thereby enhancing multimodal representation learning.

To summarize, our main contributions are threefold:
(i) We propose VIRAL, a Vision–Language Information-aware Recommendation framework that leverages VLMs to generate semantically enriched visual features;
(ii) We design an adaptive information-aware fusion module based on PID to balance complementary and redundant modality signals;
(iii) Extensive experiments on three Amazon datasets show that VIRAL achieves superior performance, improved visual utilization, and enhanced interpretability over SOTA multimodal baselines.


\begin{figure}[t!]
  \centering
\includegraphics[width=0.9\linewidth]{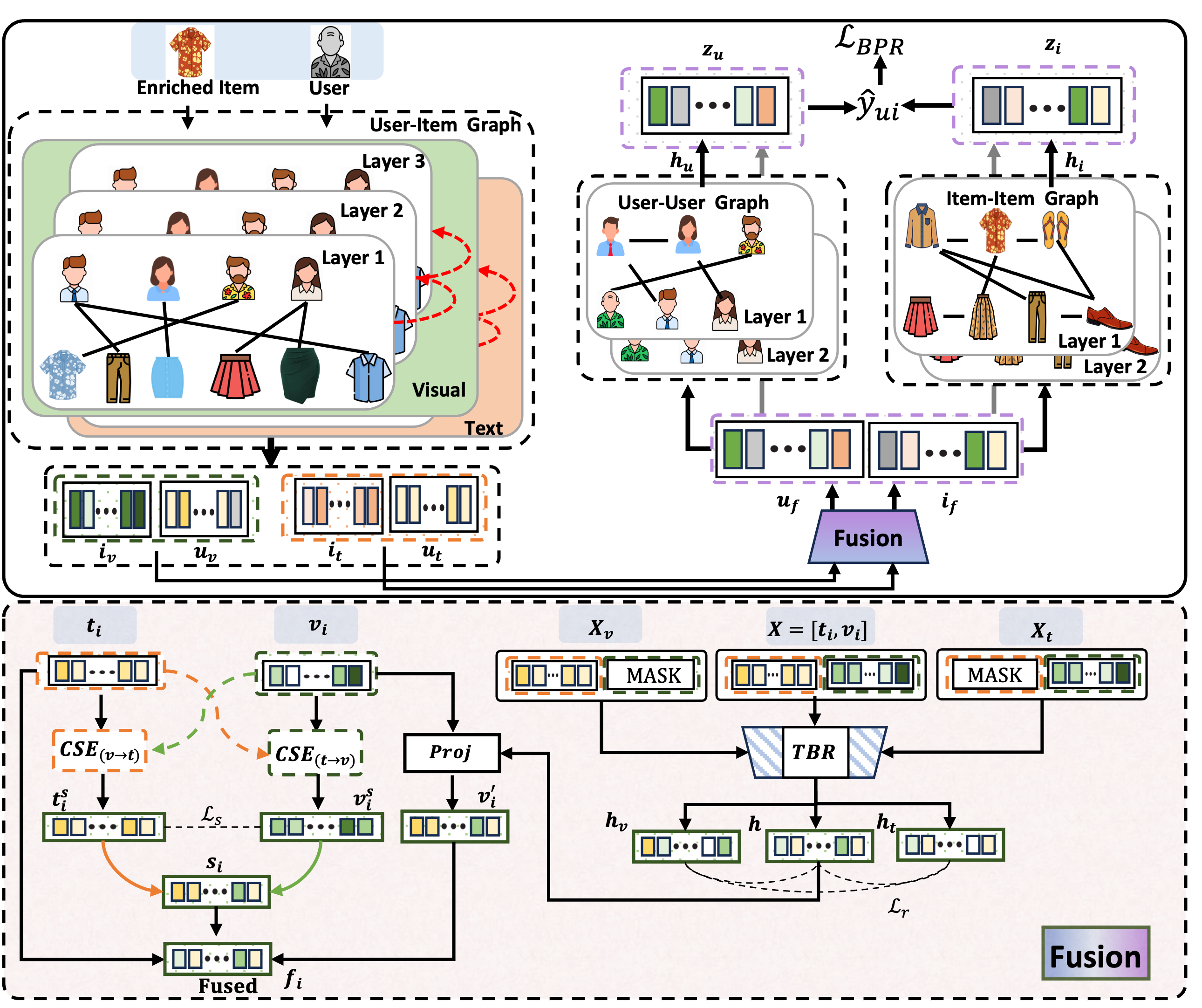}
  \vspace{-0.1in}
  \caption{Our \myname pipeline}
  \label{fig:pipeline}
\vspace{-0.25in}
\end{figure}

\section{Methodology}
\subsection{Problem and Solution Overview} 

\noindent\textbf{Problem Formulation.} We formulate the multimodal recommendation problem as predicting a user’s preference for items by jointly modeling behavioral and multimodal content signals.
Let $\mathcal{U}$ and $\mathcal{I}$ denote the sets of users and items, respectively. Each user $u \in \mathcal{U}$ interacts with a subset of items $\mathcal{I}_u \subseteq \mathcal{I}$, forming an implicit feedback matrix $\mathbf{R} \in {0,1}^{|\mathcal{U}|\times|\mathcal{I}|}$. Each item $i \in \mathcal{I}$ is associated with a visual feature $\mathbf{v}_i$ and a textual feature $\mathbf{t}_i$, representing its visual and textual modalities. The goal is to learn a function $f(u, i)$ that estimates the likelihood of user $u$ interacting with item $i$ by effectively leveraging both behavioral and multimodal information.


\noindent\textbf{Framework Overview.} As illustrated in Figure~\ref{fig:pipeline}, our proposed \myname framework operates in three main stages, each detailed in the following sections. First, adopts a graph-based architecture that captures both heterogeneous and homogeneous relations to model multimodal user–item interactions (Section~\ref{sec:graph}). Next, a VLM-based Visual Enrichment module (Section~\ref{sec:vlm}) leverages VLM guided by item titles to generate fine-grained, semantically aligned visual representations, mitigating the noise of raw image features. Finally, an Information-Aware Fusion module (Section~\ref{sec:fusion}) inspired by PID disentangles redundant, synergistic, and unique modality information to produce unified multimodal embeddings for recommendation.



\subsection{Graph-based Multimodal Interaction Modeling}
\label{sec:graph}

To capture both behavioral and semantic dependencies in multimodal recommendation, \myname constructs three complementary graphs: \textit{a heterogeneous user–item interaction graph} that models explicit behavioral relations, and two \textit{homogeneous graphs} item–item and user–user to capture semantic correlations among items and users, respectively.


\noindent\textbf{User--Item Graph.}
We take user embeddings and modality-specific item features as input and construct two user--item graphs  
$\mathcal{G} = \{\mathcal{G}_m \mid g_v, g_t\}$ to model modality-dependent interactions. 
Each graph $\mathcal{G}_m$ maintains an identical topology while preserving only the node features of modality $m$. 

Formally, the user and item representations at the $l$-th graph convolution layer are updated as:
\begin{equation}
\mathbf{u}_m^{(l)} = \sum_{i \in \mathcal{N}_u} 
\frac{1}{\sqrt{|\mathcal{N}_u|}\sqrt{|\mathcal{N}_i|}} \mathbf{i}_m^{(l-1)}, \quad
\mathbf{i}_m^{(l)} = \sum_{u \in \mathcal{N}_i} 
\frac{1}{\sqrt{|\mathcal{N}_u|}\sqrt{|\mathcal{N}_i|}} \mathbf{u}_m^{(l-1)}.
\end{equation}
Here, $\mathcal{N}_u$ and $\mathcal{N}_i$ denote the neighbor sets of user $u$ and item $i$ in $\mathcal{G}$, respectively. 
The initial user embedding $\mathbf{u}_m^{(0)}$ is randomly initialized, while the initial item embedding $\mathbf{i}_m^{(0)}$ is derived from its corresponding visual feature $\mathbf{v}_i$ or textual feature $\mathbf{t}_i$. 
The final embedding for each modality is obtained by aggregating features across all $L$ layers:
\begin{equation}
\bar{\mathbf{u}}_{m} = \sum_{l=0}^{L} \mathbf{u}_{m}^{(l)}, \quad 
\bar{\mathbf{i}}_{m} = \sum_{l=0}^{L} \mathbf{i}_{m}^{(l)}.
\end{equation}
Each modality-specific user and item embedding is then fed into the \textit{Information-Aware Fusion module} (Section~\ref{sec:fusion}) 
to obtain unified representations $\mathbf{u}_f$ and $\mathbf{i}_f$, 
which are subsequently propagated through the homogeneous graphs to capture user, item respectively.

\noindent\textbf{Item--Item Graph.}
To model semantic correlations between items, we construct an item--item homogeneous graph $\mathcal{G}_i$ via a $k$-nearest neighbor (KNN) strategy using item features from each modality $m$. 
The similarity between items $(i, i') \in \mathcal{I}$ is measured by cosine similarity, and the final similarity score is defined as 
$S_{i,i'} = \tfrac{1}{2}(S_{i,i'}^{v} + S_{i,i'}^{t})$, 
combining visual and textual similarities. 
Only the top-$k$ neighbors are retained, while the remaining edges are pruned. 
The item representation is updated as:
\begin{equation}
\mathbf{h}_i^{(l)} = \sum_{i' \in \mathcal{N}_i} S_{i,i'} \mathbf{h}_{i'}^{(l-1)},
\label{item-item}
\end{equation}
where $\mathcal{N}_i$ denotes the neighbor set of item $i$ in $\mathcal{G}_i$ and 
$\mathbf{h}_i^{(0)}$ is initialized with $\mathbf{i}_f$. 
For the \textbf{User--User Graph}, edges are constructed based on the number of co-interacted items, and only the top-$k$ neighbors are preserved. 
User representations are updated analogously to the aggregation process in Equation~(\ref{item-item}). 
The final user and item representations, $\mathbf{z}_u$ and $\mathbf{z}_i$, are obtained by integrating the homogeneous graph embeddings with the residual fusion signals. 
The predicted interaction score is then computed as:
\begin{equation}
\hat{y}_{ui} = \mathbf{z}_i^{\top} \mathbf{z}_u.
\end{equation}

\subsection{VLM-based Visual Enrichment}
\label{sec:vlm}
We leverage vision–language models (VLMs) to transform visual inputs into semantically enriched textual representations through task-specific prompting. 
Given a product image $\mathbf{v}_i$ and its title $\mathbf{t}^{title}_i$, the prompt $\mathcal{P}(\mathbf{v}_i, \mathbf{t}^{title}_i)$ integrates both modalities to guide the VLM toward generating item-relevant descriptions. 
\begin{figure}[h!]
  \centering
\includegraphics[width=1\linewidth]{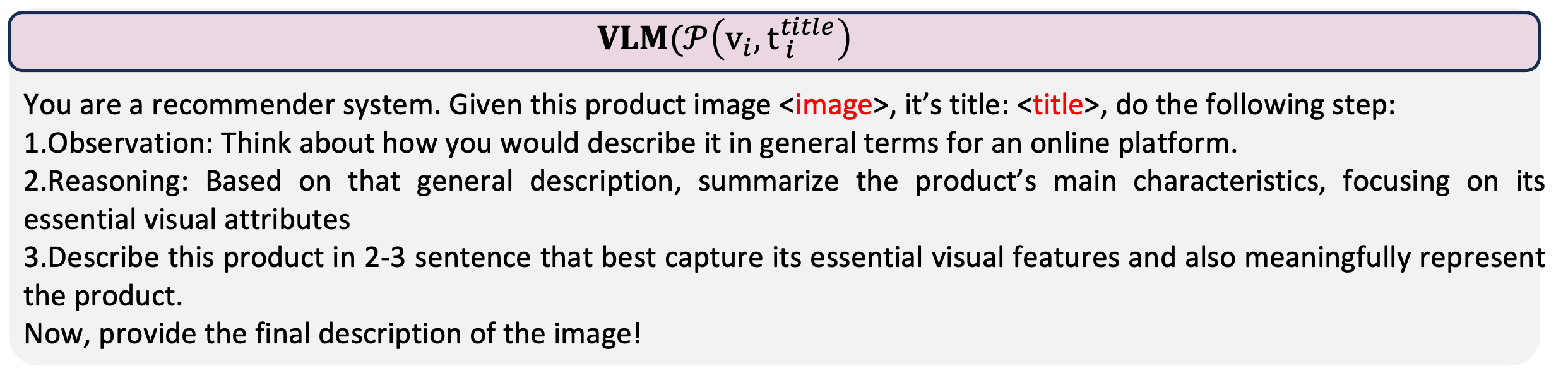}
  \vspace{-0.3in}
  \caption{VLM enriches visual features}
  \label{fig:vlm}
\vspace{-0.2in}
\end{figure}

To improve semantic reasoning and accuracy, we employ a Chain-of-Thought (CoT) prompting strategy that encourages step-by-step generation of enriched item descriptions. 
As illustrated in Figure~\ref{fig:vlm}, the VLM produces descriptive text, which is then encoded into a dense embedding serving as the enhanced visual representation:
\begin{equation}
    \mathbf{v}_i = \mathrm{Enc}\big(\mathrm{VLM}(\mathcal{P}(\mathbf{v}_i, \mathbf{t}^{title}_i))\big),
\end{equation}
where $\mathrm{Enc}(\cdot)$ denotes a text encoder that maps generated descriptions into the feature space.


\subsection{Information-Aware Fusion}
\label{sec:fusion}
\noindent
From the information-theoretic perspective, the relationship of multimodal inputs $X_i=\{v_i, t_i\}$ and target $Y$ can be decomposed as:
\begin{equation}
\label{pid}
I(X_1, X_2; Y) = R + S + U_v + U_t,    
\end{equation}
where $R$ denotes the redundant information shared by both modalities, $S$ the synergistic information that emerges only through their joint interaction, and $U_v, U_t$ the unique contributions of the visual and textual modalities, respectively. 
To realize this decomposition in representation space, we explicitly estimate $S$ and $R$, and derive the unique components based on these estimates.

\noindent\textbf{Cross-Modal Synergy Estimation (CSE).} 
We first utilize a cross-modal Transformer to model interactions between modalities:
\begin{equation}
    \mathbf{h}_{v \rightarrow t} = \mathrm{Trans}(\mathbf{v}, \mathbf{t}),
\end{equation}
where $\mathbf{h}_{v \rightarrow t}$ represents attended features obtained by querying vision over text (and symmetrically $\mathbf{h}_{t \rightarrow v}$ for text over vision). 
The shared information between these two attention directions constitutes the synergistic signal, which we estimate as:
\begin{equation}
    \mathbf{s} = \tfrac{1}{2}(\mathbf{h}_{v \rightarrow t} + \mathbf{h}_{t \rightarrow v}).
\end{equation}

\noindent\textbf{Transformer-based Redundancy Estimation (TBR).}
To extract shared (redundant) features $R$, we construct three input variants:
\begin{equation}
    \mathbf{X} = [\mathbf{v}, \mathbf{t}], \quad 
    \mathbf{X}_v = [\mathbf{v}, \texttt{[MASK]}_t], \quad 
    \mathbf{X}_t = [\texttt{[MASK]}_v, \mathbf{t}],
\end{equation}
where $\texttt{[MASK]}_t$ and $\texttt{[MASK]}_v$ are special tokens used to remove a modality. 
Each variant is passed through a shared Transformer encoder $F(\cdot)$ followed by a projection layer $g(\cdot)$:
\begin{equation}
    \mathbf{h} = g(F(\mathbf{X})), \quad
    \mathbf{h}_v = g(F(\mathbf{X}_v)), \quad
    \mathbf{h}_t = g(F(\mathbf{X}_t)).
\end{equation}
Aligning the unimodal encodings $\mathbf{h}_v$ and $\mathbf{h}_t$ with the full multimodal encoding $\mathbf{h}$ allows the model to capture modality-invariant information, which we define as the redundant feature:
\begin{equation}
    \mathbf{r} = \tfrac{1}{3}(\mathbf{h} + \mathbf{h}_v + \mathbf{h}_t).
\end{equation}

\noindent\textbf{Fusion.}
To obtain the unique visual contribution $U_v$, we remove the redundant component $\mathbf{r}$ from the visual representation $\mathbf{v}$ by orthogonal projection:
\begin{equation}
    \mathbf{v}' = \mathbf{v} - \text{Proj}_{\mathbf{r}}(\mathbf{v}), \quad
    \text{Proj}_{\mathbf{r}}(\mathbf{v}) = 
    \frac{\mathbf{v}^\top \mathbf{r}}{\|\mathbf{r}\|_2^2}\mathbf{r}.
\end{equation}
This yields $\mathbf{v}'$, the unique and redundancy-free visual representation orthogonal to $\mathbf{r}$. 
The final item representation concatenates three components—textual, synergistic, and unique visual features—as:
\begin{equation}
    \mathbf{i}_f = [\, \mathbf{t} \; || \; \mathbf{s} \; || \; \mathbf{v}' \,],
\end{equation}
where $||$ denotes concatenation. 
To model user-specific modality preferences, we employ a learnable weight vector $\textbf{u}^w = [u^w_0, u^w_1, u^w_2]$ derived from user IDs, producing a fused user embedding:
\begin{equation}
    \mathbf{u}_f = [\, u^w_0 \mathbf{t} \; || \; u^w_1 \mathbf{s} \; || \; u^w_2 \mathbf{v}' \,].
\end{equation}

\subsection{Optimization}
\label{sec:op}

To regulate modality interactions, we employ \textbf{InfoNCE} that explicitly models the synergy information shared between modalities. 
\begin{equation}
    \mathcal{L}_{\text{s}} 
    = 
        \mathcal{L}_{\text{InfoNCE}(h_{v \rightarrow t}, h_{t \rightarrow v})} 
\end{equation}
where $\mathcal{L}_{\text{InfoNCE}}$ is the standard contrastive loss that promotes alignment between 2 presentations. Similarly, we align the modality's common information, through the three representations:
\begin{equation}
    \mathcal{L}_{\text{r}} =
    \mathcal{L}_{\text{InfoNCE}}(h, h_v) +
    \mathcal{L}_{\text{InfoNCE}}(h, h_t) +
    \mathcal{L}_{\text{InfoNCE}}(h_v, h_t),
\end{equation}
which maximizes the mutual information among $(h, h_v, h_t)$ to learn a modality redundant feature \( \mathbf{r} \). The final training objective is:
\begin{equation}
    \mathcal{L} 
    = \mathcal{L}_{\text{rec}} 
    + \lambda (\mathcal{L}_{\text{s}}+  \mathcal{L}_{\text{r}}).
\end{equation}
where $\mathcal{L}_{\text{rec}}$ is the recommendation BPR loss and $\lambda$ is a hyperparameter controlling the strength of fusion regularization.
\section{Experiments}
\subsection{Experimental Settings}


\begin{table*}[ht!]
\centering
\resizebox{0.9\textwidth}{!}{%
\begin{tabular}{l|l|cc|ccccccc|ccc|c}
\hline
\multirow{2}{*}{\textbf{Dataset}} 
& \multirow{2}{*}{\textbf{Metric}} 
& BPR & LightGCN & VBPR & MMGCN 
  & GRCN & SLMRec & LATTICE 
  & FREEDOM & DRAGON & MENTOR & COHESION & \myname 
  & \multirow{2}{*}{\textbf{Improv.}} \\ 
\cline{3-14}
& & \cellcolor[HTML]{E0E0E0}{UAI-12} & \cellcolor[HTML]{E0E0E0}{SIGIR-20} & \cellcolor[HTML]{E0E0E0}{AAAI-16} & \cellcolor[HTML]{E0E0E0}{MM-19} & \cellcolor[HTML]{E0E0E0}{MM-20} & \cellcolor[HTML]{E0E0E0}{2021} & \cellcolor[HTML]{E0E0E0}{MM-21} & \cellcolor[HTML]{E0E0E0}{MM-23} & \cellcolor[HTML]{E0E0E0}{ECAI-23} & \cellcolor[HTML]{E0E0E0}{AAAI-25} & \cellcolor[HTML]{E0E0E0}{SIGIR-25} &
\cellcolor[HTML]{E0E0E0}{2025}&  \\ 
\hline

\multirow{3}{*}{Baby} 
& R@10  & 0.0357 & 0.0479 & 0.0423 & 0.0421 & 0.0532  & 0.0521 & 0.0547 & 0.0627 & 0.0513 & 0.0678 & \underline{0.0680} & \textbf{0.0705} & 3.7\% \\
& R@20  & 0.0575 & 0.0754 & 0.0663 & 0.0660 & 0.0824  & 0.0772 & 0.0850 & 0.0992 & 0.0803 & 0.1048 & \underline{0.1052} & \textbf{0.1083} & 2.9\% \\
& N@10  & 0.0192 & 0.0257 & 0.0223 & 0.0220 & 0.0282  & 0.0289 & 0.0292 & 0.0330 & 0.0278 & \underline{0.0362} & 0.0354 & \textbf{0.0378} & 4.4\% \\
& N@20  & 0.0249 & 0.0328 & 0.0284 & 0.0282 & 0.0358  & 0.0354 & 0.0370 & 0.0424 & 0.0352 & 0.0450 & \underline{0.0454} & \textbf{0.0474} & 4.4\% \\

\hline

\multirow{3}{*}{Sports} 
& R@10  & 0.0432 & 0.0569 & 0.0558 & 0.0401 & 0.0599  & 0.0663 & 0.0620 & 0.0717 & 0.0588 & \underline{0.0763} & 0.0752 & \textbf{0.0788} & 3.3\% \\
& R@20  & 0.0653 & 0.0864 & 0.0856 & 0.0636 & 0.0919  & 0.0990 & 0.0953 & 0.1089 & 0.0899 & \underline{0.1139} & 0.1137 & \textbf{0.1168} & 2.5\% \\
& N@10  & 0.0241 & 0.0311 & 0.0307 & 0.0209 & 0.0330  & 0.0365 & 0.0335 & 0.0385 & 0.0324 & \underline{0.0409} & 0.0409 & \textbf{0.0430} & 5.1\% \\
& N@20  & 0.0298 & 0.0387 & 0.0384 & 0.0270 & 0.0413  & 0.0450 & 0.0421 & 0.0481 & 0.0404 & \underline{0.0511} & 0.0503 & \textbf{0.0528} & 3.3\% \\

\hline

\multirow{3}{*}{Clothing} 
& R@10  & 0.0206 & 0.0361 & 0.0281 & 0.0227 & 0.0421  & 0.0442 & 0.0492 & 0.0629 & 0.0452 & \underline{0.0668} & 0.0665 & \textbf{0.0681} & 1.9 \% \\
& R@20  & 0.0303 & 0.0544 & 0.0415 & 0.0361 & 0.0657  & 0.0659 & 0.0733 & 0.0941 & 0.0675 & \underline{0.0989} & 0.0983 & \textbf{0.1013} & 2.4\% \\
& N@10  & 0.0114 & 0.0197 & 0.0158 & 0.0120 & 0.0224  & 0.0241 & 0.0268 & 0.0341 & 0.0242 & \underline{0.0360} &  0.0358 & \textbf{0.0374}  & 3.9\% \\
& N@20  & 0.0138 & 0.0243 & 0.0192 & 0.0154 & 0.0284  & 0.0296 & 0.0330 & 0.0420 & 0.0298 & \underline{0.0441} & 0.0438 & \textbf{0.0462} & 4.8\% \\

\hline
\end{tabular}%
}
\caption{Overall performance.  Bold indicates best performance in each row.}
\label{tab:main}
\vspace{-0.2in}
\end{table*}


\begin{table}[ht!]
\centering
\resizebox{0.9\linewidth}{!}{%
\begin{tabular}{c|l|ccc|ccc}
\hline
\multirow{2}{*}{\textbf{Module}} & 
\multirow{2}{*}{\textbf{Approach}} & 
\multicolumn{3}{c|}{\textbf{Baby}} & 
\multicolumn{3}{c}{\textbf{Sports}} \\
\cline{3-5} \cline{6-8}
 &  & R@10 & R@20 & N@20 & R@10 & R@20 & N@20 \\
\hline
\multirow{2}{*}{\shortstack{Image \\  Feature}} 
 & Original   & 0.0668 & 0.1026 & 0.0450 & 0.0751 & 0.1126 & 0.0510\\
 & VLM w.o title & 0.0692 & 0.1077 & 0.0469 & 0.0770 & 0.01159 & 0.0515\\
\hline
\multirow{3}{*}{\shortstack{Multimodal \\  Fusion}}
 & Pooling    & 0.0668 & 0.1031 & 0.0452 & 0.0758 & 0.1130 & 0.0506\\
 & Concat     & 0.0677 & 0.1035 & 0.0454 & 0.0778 & 0.1142 & 0.0511\\
 & Weighted Concat  & 0.0668 & 0.1068 & 0.0461 & 0.0770 & 0.1160 & 0.0520\\
\hline
\multicolumn{2}{c}{\textbf{\myname}}
 & \textbf{0.0705} & \textbf{0.1083} & \textbf{0.474} & \textbf{0.0788} & \textbf{0.1168} & \textbf{0.0528}\\
\hline
\end{tabular}%
}
\caption{Ablation study. Contribution of each sub-module.}
\label{tab:ablation}
\vspace{-0.4in}
\end{table}

\noindent \textbf{Dataset. }Following prior studies~\cite{Mentor, FREEDOM}, we conduct experiments on three categories of the Amazon Review dataset\footnote{Publicly available at \url{http://jmcauley.ucsd.edu/data/amazon/links.html}}
: \textit{Baby}, \textit{Sports and Outdoors}, and \textit{Clothing, Shoes and Jewelry}, denoted as \textit{Baby}, \textit{Sports}, and \textit{Clothing} for brevity. We adopt the standard 5-core setting, ensuring that each user and item has at least five interactions. 

\noindent \textbf{Evaluation. }
We compare \myname with the following SOTA methods:
BPR, LightGCN~\cite{LightGCN}, VBPR~\cite{VBPR}, MMGCN~\cite{MMGCN}, {GRCN}~\cite{RGCN}, {SLMRec}~\cite{SLMREC}, {LATTICE}~\cite{LATTICE}, {FREEDOM}~\cite{FREEDOM}, DRAGON~\cite{DRAGON},{MENTOR}~\cite{Mentor}, {COHESION}~\cite{COHESION}. Performances are evaluated on 2 metrics: Recall@$10$ (R@$10$) and NDCG@$10$ (N@$10$). We follow the setting of ~\cite {FREEDOM} with a random data splitting of 8:1:1 for training, validation, and testing. 

\noindent \textbf{Implementation Details.} 
We implement \textsc{\myname} with Xavier initialization and a batch size of 1024, learning rate of 0.001, embedding dimension \(d = 64\) and two layers for user--item, item--item, and user--user graphs,  with top-\(k = 10\) and \(\lambda = 0.1\).  Image descriptions are generated using the {Qwen-2.5-VL-7B} model.  Experiments are conducted on an {NVIDIA A5000 (24GB)} GPU. VLM-generated descriptions and textual content (concatenated from item’s brand, title, description, and category) are both encoded using Sentence-BERT\footnote{https://huggingface.co/sentence-transformers}. 

\subsection{Experimental Results}
\noindent \textbf{Overall Performance.} As shown in Table~\ref{tab:main}, our \myname\ consistently outperforms the two strongest baselines, \textit{MENTOR} and \textit{COHESION}, by a notable margin across all datasets and evaluation metrics. This demonstrates that integrating features derived from VLMs with our information-aware fusion module yields superior performance compared to using raw features or conventional fusion strategies, providing an effective solution for multimodal representation learning. When compared with \cite{DRAGON}, which adopts a similar multimodal encoder, replacing the original visual features with VLM-based representations and our fusion strategy significantly improves performance.

\begin{figure}[t!]
    \centering
        \includegraphics[width=0.9\linewidth]{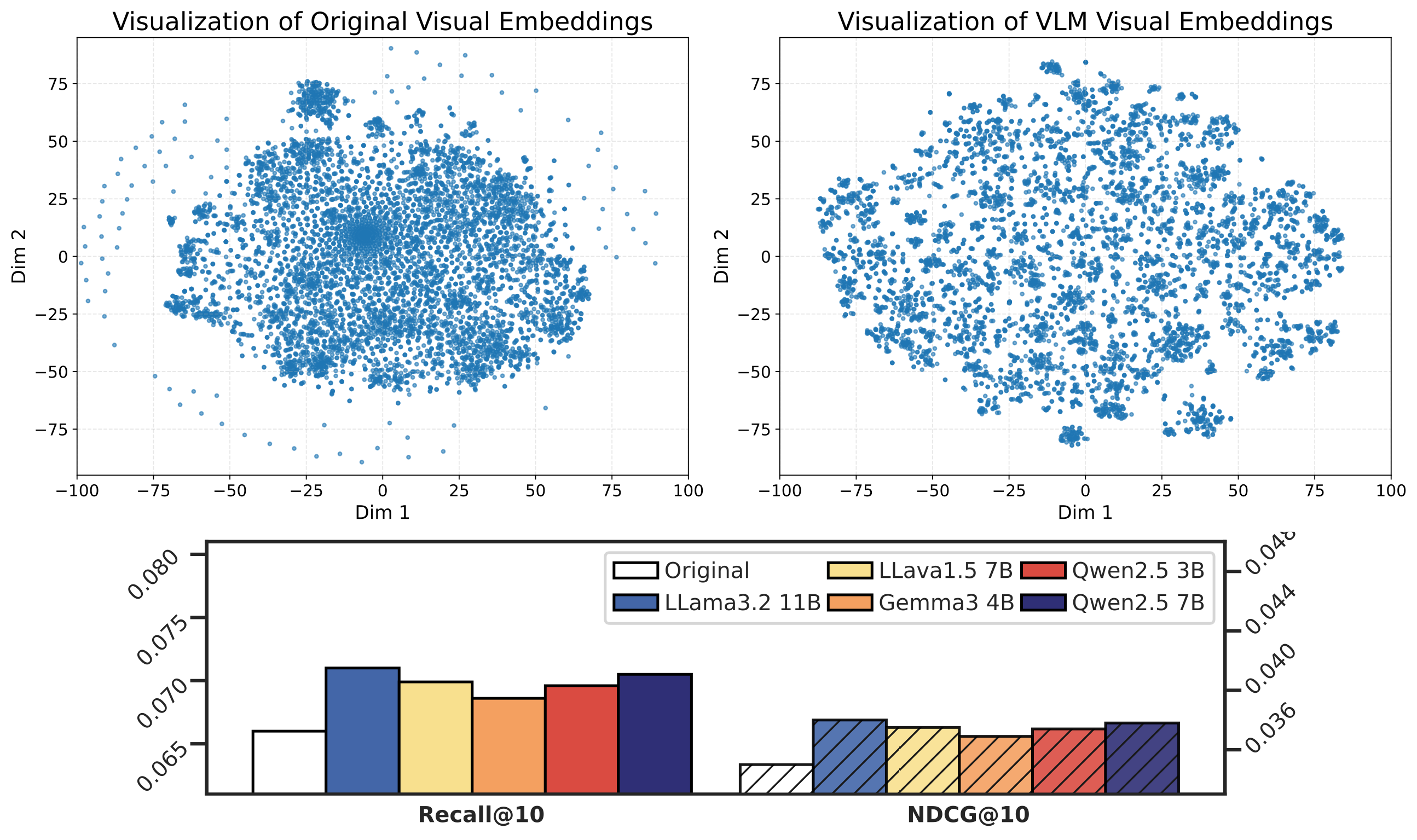}
    \vspace{-0.1in}
    \caption{
Experiments on Baby. [Top] Item embedding visualization. 
[Bottom] Performance of VLM models.}
\label{fig:distribution}
    \vspace{-0.2in}
\end{figure}

\noindent \textbf{Ablation.} Furthermore, we investigate the contribution of each proposed module.   Table~\ref{tab:ablation} reports the performance when removing either the VLM-based feature generation or the information-aware fusion. The performance degradation observed highlights the necessity of both components. In particular, for image features, the original visual embeddings yield the lowest accuracy, followed by plain VLM features, while our task-specific VLM guided by item titles (\myname) achieves the best performance. Likewise, alternative fusion strategies perform worse than our proposed information-aware fusion.  Moreover, Figure~\ref{fig:modality} illustrates that incorporating VLM features balances the contribution of each modality, preventing over-reliance on textual signals. Finally, in Figure \ref{fig:distribution}, we also evaluate \myname\ with different VLM backbones, and the results confirm that our framework maintains stable and consistent improvements across models, demonstrating its general applicability. Furthermore, we visualize the original and VLM-based visual embeddings. The original features are scattered with many outliers, while the VLM-based embeddings show a more structured distribution, indicating richer and more semantically aligned representations.

\vspace{-0.1in}
\section{Conclusion}

In this work, we propose \myname, a multimodal recommendation framework that leverages VLM to generate fine-grained item representations and an information-aware fusion module to integrate modalities effectively. Experiments on three Amazon product datasets demonstrate that \myname\ consistently outperforms strong multimodal baselines and significantly strengthens the contribution of the visual modality. We plan to further enhance the framework by jointly training the VLM with recommendation signals, enabling task-adaptive visual–textual understanding for recommendation.

\bibliographystyle{ACM-Reference-Format}
\bibliography{sample-base}

\appendix

\end{document}